\documentclass[aps,prl,twocolumn,showpacs]{revtex4}
\usepackage{amssymb}
\usepackage{amsmath}
\usepackage{graphicx}
\usepackage{upgreek}

\begin{document}
\title{Second-order photon correlation measurement with picosecond resolution}

\author{Aymeric Delteil$^{1,2}$}   
\author{Chun Tat Ngai$^{1,3}$ }     
\author{Thomas Fink$^{1}$}       
\author{Ata\c{c} \.Imamo\u{g}lu$^{1}$}   

\affiliation{1 Institute of Quantum Electronics, ETH Zurich, CH-8093
Zurich, Switzerland. \\
2 Groupe d'\'Etude de la Mati\`ere Condens\'ee, Universit\'e de Versailles Saint-Quentin-en-Yvelines, CNRS UMR 8635, Universit\'e Paris-Saclay, 45 Avenue des \'Etats-Unis, 78035 Versailles cedex, France \\ 
3 Department of Physics, University of Basel, Klingelbergstrasse 82, 4056 Basel, Switzerland
}


\begin{abstract}

The second-order correlation function of light $g^{(2)}(\tau)$ constitutes a pivotal tool to quantify the quantum behavior of an emitter and in turn its potential for quantum information applications. The experimentally accessible time resolution of $g^{(2)}(\tau)$ is usually limited by the jitter of available single photon detectors.
Here, we present a versatile technique allowing to measure $g^{(2)}(\tau)$ from a large variety of light signals with a time resolution given by the pulse length of a mode-locked laser. The technique is based on frequency upconversion in a nonlinear waveguide, and we analyze its properties and limitations by modeling the pulse propagation and the frequency conversion process. We measure $g^{(2)}(\tau)$ from various signals including light from a quantum emitter – a confined exciton-polariton structure -- revealing its quantum signatures at a scale of a few picoseconds and demonstrating the capability of the technique.

\end{abstract}

\maketitle
The signature of single photon emission is a vanishing second-order correlation function $g^{(2)}(\tau)$ at zero delay ($\tau = 0$). Beyond being one of the most remarkable non-classical properties of light emitted by quantum systems, such behavior also turns out to be an essential resource for optical quantum information technologies. Vanishing $g^{(2)}(0)$ has been consistently observed in emission from a wide variety of physical systems, from single atoms to solid state emitters~\cite{Grangier86, Lounis00, Aharonovich16}, and is typically measured by recording a histogram of the time delays between pairs of photon detection events. In continuous-wave (cw) emission, the timescale at which $g^{(2)}(\tau)$ recovers classical characteristics is in most cases comparable with the lifetime of the emitter excited state. Therefore, the observation of non-classical signatures in $g^{(2)}(\tau)$ is limited to the cases where the resolution of available single-photon detectors (typically several tens to hundreds of picoseconds) is shorter than this characteristic timescale.

In order to overcome this limitation, alternative techniques can be used. Streak cameras provide time resolutions of order picosecond or lower, but their very low repetition rate ($\lesssim 1$~kHz) precludes their use for measuring two-photon correlations of weak signals~\cite{Ueda88,Bayer10}. On another note, frequency conversion using a short laser pulse can also provide high time resolution~\cite{Mahr75}. This technique has been used to characterize the fluorescence decay of a wide range of physical systems under pulsed excitation, with  a resolution below a picosecond~\cite{Block86,Shah88,Chesnoy89,Markovitsi15}. In these experiments, the emitted light is mixed in a nonlinear crystal with a short laser pulse (termed pump pulse) that is synchronized with the excitation laser. Since frequency conversion occurs only during the pump pulse, the latter serves as a fast gate that provides accurate information about the emission time. The converted photons are then detected by standard detectors, and the signal envelope is reconstructed by recording the detection rate while varying the time delay between the signal and pump pulses. Recent progress in the conversion efficiency of nonlinear crystals based on QPM has allowed time-resolved detection of single-photons generated by spontaneous parametric downconversion~\cite{Kuzucu08} and realization of fast gating of quantum dot single-photon emission~\cite{Rakher11,DeGreve12,Yu15} based on the same principle. 

Here, we extend the principle of these experiments to the measurement of the second-order correlation function of arbitrary light signals that need no synchronization with the pump laser. In particular, our method is well suited for measuring  $g^{(2)}(\tau)$ of continuous wave emission. The nonlinear medium we use is a periodically poled lithium niobiate (PPLN) crystal waveguide with intrinsically high conversion efficiency, allowing for measurement of very weak signals down to the single photon level. Thanks to the waveguide configuration, the QPM condition can be fulfilled by simply tuning the wavelength of the pump laser. Consequently, a wide range of signal wavelengths can be measured using the same set-up. Our technique is expected to play a crucial role for characterizing the purity and indistinguishability of ultrafast single photon sources with a radiative decay rate exceeding the bandwidth of single photon detectors, based for instance on confined exciton-polaritons~\cite{Munoz19,Delteil19} or single emitters in microcavities and plasmonic structures~\cite{Bayer01,Mikkelsen16}.

\begin{figure}[htbp]
\centering
\includegraphics[width=\linewidth]{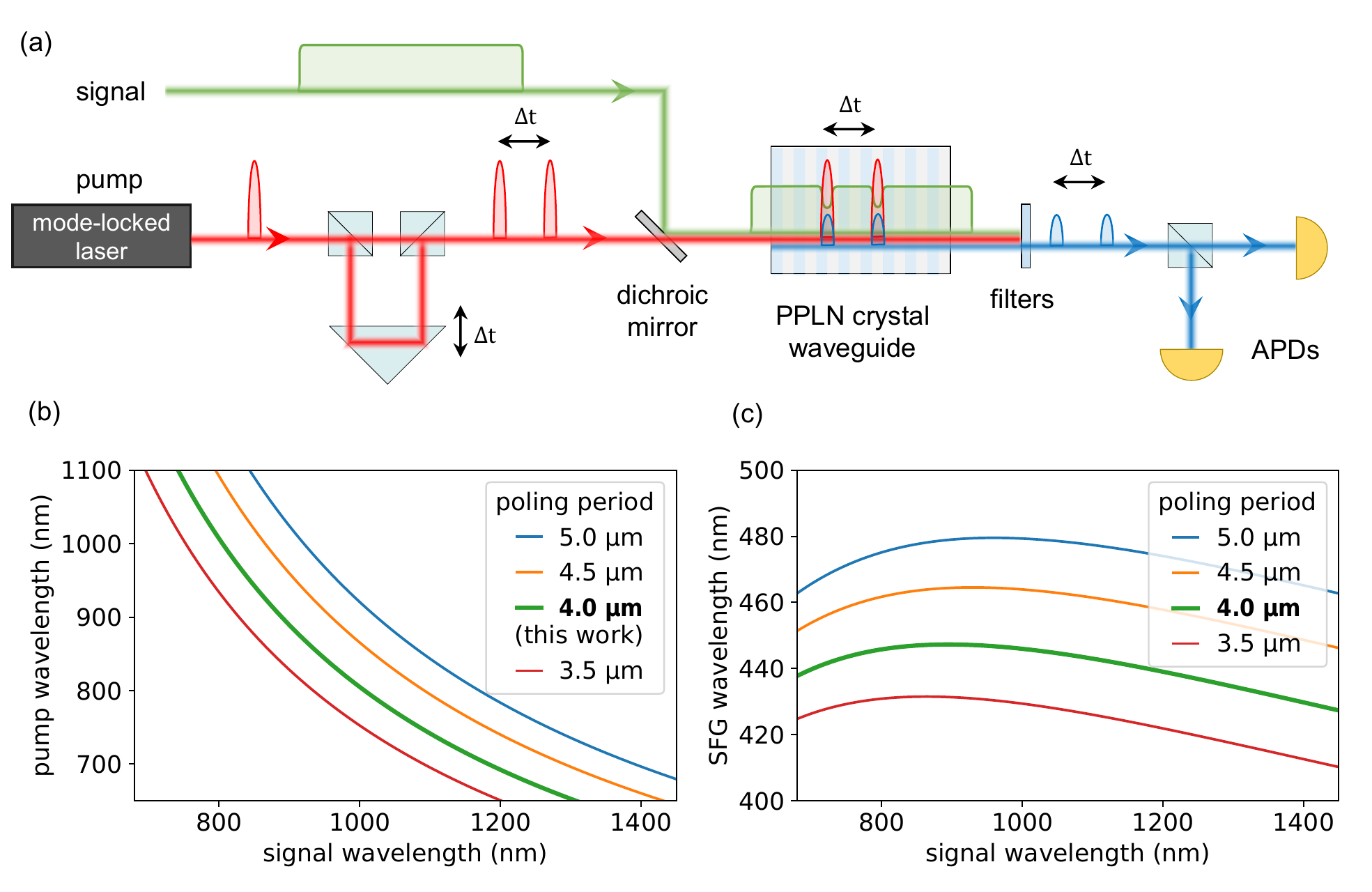}
\caption{(a) Experimental set-up. Two ps-pulses (red beam) with variable delay are mixed with an input signal (green beam) in a PPLN crystal waveguide. The upconverted pulses (blue beam) are detected in a Hanbury Brown and Twiss setup. (b) Pump wavelength as a function of the signal wavelength calculated from the phase-matching relations in PPLN with various poling period. (c) Corresponding final wavelength as a function of the signal wavelength.}
\label{fig:1}
\end{figure}

The setup is depicted in figure~\ref{fig:1}a. It is based on a Ti:sapphire (pump) laser of pulse length 2.5~ps and repetition rate 76~MHz from which we generate pairs of pulses separated by a variable delay. A dichroic mirror combines the pump pulses and the input signal into a single spatial mode, which is focused on the input facet of a commercial PPLN waveguide (from HC Photonics) of poling period 3.96~$\upmu$m, length 12.5~mm and mode size $4~\upmu$m$\times 6~\upmu$m. The polarization of both the pump and the signal are made parallel to the extraordinary axis of the PPLN crystal. Provided that the pump wavelength is tuned to be in quasi-phase matching (QPM) condition with the signal wavelength, sum frequency generation (SFG) occurs in the PPLN waveguide (figure~\ref{fig:1}b). The tunability range of our pump laser allows QPM for signal wavelength from 750~nm to 1150~nm and could be extended further by choosing a different poling period of the PPLN crystal (figure~\ref{fig:1}b). The upconverted signal wavelength depends only weakly on the signal wavelength and is about 445~nm (figure~\ref{fig:1}c).
After filtering out the signal, pump pulse, and its second harmonic from the output of the waveguide, the upconverted pulses are collected in the input port of a fiber beamsplitter. The two output ports are coupled to standard avalanche photodiodes (APDs) of time resolution $\sim300$~ps. The coincidence rate $C(\Delta t)$ is recorded as a function of the time delay $\Delta t$ between the pulses. $C(\Delta t)$ is then normalized by the average coincidence rate between pairs of photons originating from different (uncorrelated) pump pulse periods, which yields $c(\Delta t) = C(\Delta t)/ \left\langle C(\Delta t + nT) \right\rangle _{n \neq 0} $, where $T$ is the repetition rate of the pump laser. Note that $C(\Delta t)$ contains equal contributions from photon pairs converted by the same pump pulse and by the two different pump pulses. Therefore, the resulting normalized coincidence rate reads $c(\Delta t) = \dfrac{1}{2} \left(g^{(2)}(\Delta t) + g^{(2)}(0)\right)$. The second order correlation function can then be retrieved straightforwardly as: $g^{(2)}(\Delta t) = 2c(\Delta t) - c(0)$.

\begin{figure}[htbp]
\centering
\includegraphics[width=\linewidth]{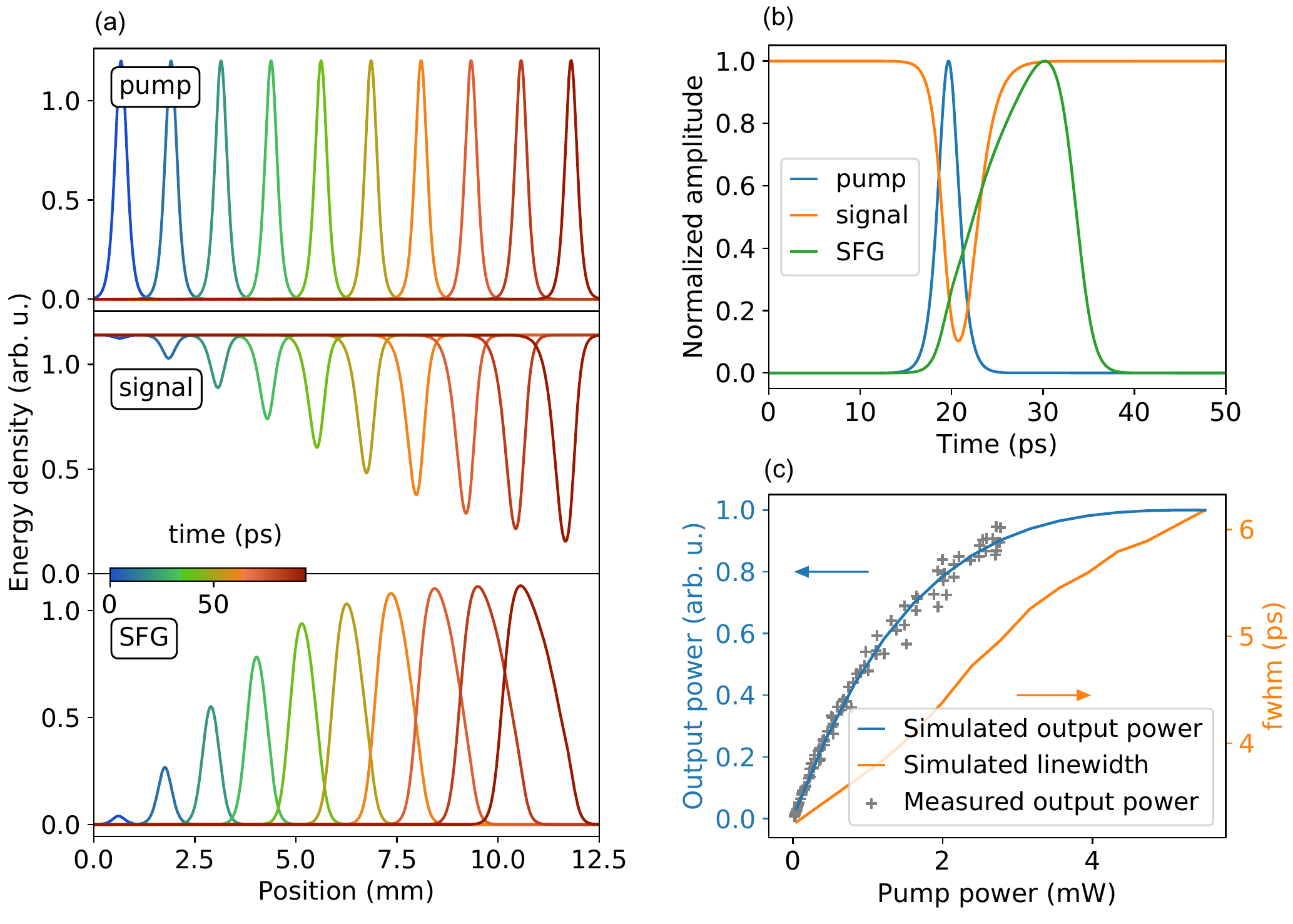}
\caption{(a) Spatial envelope of the pump (upper panel), the signal (middle panel), and the SFG (lower panel) at various propagation times, indicated by the color scale. The span of the $x$ axis is equal to the waveguide length. (b) Normalized temporal envelope of the three output pulses. The FWHM of the dip in the signal envelope provides the time resolution of our setup. (c) Continuous curves: calculated output power (blue curve) and time resolution (orange curve) as a function of the pump power. Crosses: measured output power as a function of the pump power.}
\label{fig:2}
\end{figure}

The limitations to the time resolution of the setup can be identified by simulating the pulse propagation and frequency conversion in the waveguide. We implemented a 1D propagation model that is detailed in the Supplementary Material. We model the pump pulse by a sech-squared pulse of width 2.5~ps as determined from pulse autocorrelation measurements, while the signal is initially taken as constant. Figure~\ref{fig:2}a shows the results of the simulations. The pump peak power is taken much higher than the signal power. Under this assumption, the pump pulse remains essentially undepleted as it propagates along the crystal (upper panel). The signal envelope (middle panel) exhibits a dip originating from upconverted photons. The small group velocity difference between the pump and signal ensures that this dip keeps a substantial overlap with the pump pulse during propagation. Finally, the SFG (lower panel) exhibits an increasing pulse length during the propagation due to sizably smaller group velocity. We emphasize that the time resolution is not affected by the SFG group velocity but only by the group velocity difference between the pump and the signal, which defines the range of signal emission times probed by the upconversion process. Figure~\ref{fig:2}b shows the temporal envelope of the three waves at the output of the 12.5~mm waveguide. The width of the dip in the signal envelope provides the time resolution of the setup, which we find to be 4.0~ps, limited by the pump pulse length, the group velocity difference between pump and signal, and the upconversion saturation that leads to additional broadening. Figure~\ref{fig:2}c shows the pump power dependence of the integrated upconverted signal (blue curve) and time resolution of the setup, given by the FWHM of the signal dip (orange curve). The output power exhibits a saturation behavior accompanied with a broadening of the resolution. The experimental conversion efficiency presents the same saturation behavior, allowing us to fit the saturation power (or equivalently the conversion efficiency) in the model. The final choice of the pump power results from a trade-off between conversion efficiency and time resolution. In the following, we set the pump power to be 1.5~mW, corresponding to the parameters used in the calculations of figure~\ref{fig:2}a and b.

\begin{figure}[htbp]
\centering
\includegraphics[width=\linewidth]{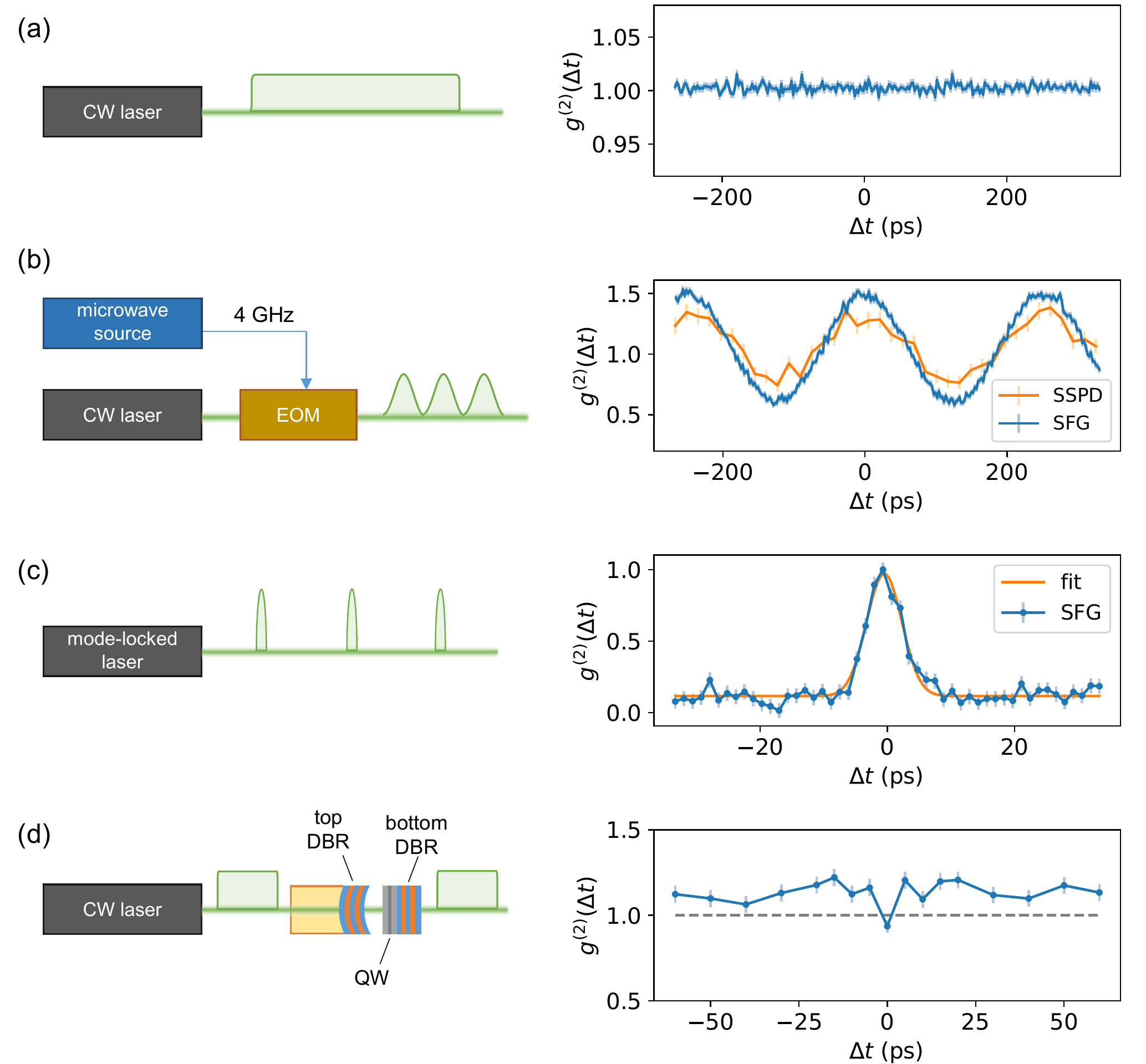}
\caption{Experimental $g^{(2)}(\Delta t)$ for various input signals. (a) CW laser, (b) CW laser modulated at 4~GHz by an EOM (blue curve). The orange curve is the $g^{(2)}(\Delta t)$ of the same signal measured with a SSPD of time resolution $\sim40$~ps. (c) ps-laser with a repetition rate incommensurable with the pump. The orange curve is a Gaussian fit to the data. (d) Light transmitted through a confined polariton structure. QW: quantum well; DBR: distributed Bragg mirror.}
\label{fig:3}
\end{figure}

Figure~\ref{fig:3} displays the measurement results with various input signals. We start with cw laser light at a wavelength of 812~nm. The pump laser is tuned to 990~nm to obtain QPM. The result is shown fig.~\ref{fig:3}a:  As expected, the measured $g^{(2)}$ is flat over the whole range, with a mean value of $1.00 \pm 0.01$, ensuring absence of specious correlations from the setup. The blue curve in fig.~\ref{fig:3}b  shows the $g^{(2)}$ of a cw laser that is modulated at 4~GHz using an electro-optical modulator (EOM). The modulation frequency is incommensurable with the pump repetition rate. The $g^{(2)}$ exhibits oscillations of visibility 38~\%, corresponding to half of the modulation depth of the EOM. We compare with the result obtained with a superconducting single-photon detector (SSPD) of resolution $\sim$40~ps per channel (including electronics) using a standard start-stop technique (orange curve in fig.~\ref{fig:3}b). In the latter case we observe similar oscillations, with reduced visibility of 26~\%, limited by the resolution of the SSPD. In order to estimate the resolution of our system, we then use the signal from a ps-laser of pulse length 2.5~ps and wavelength 812~nm as an input. The repetition rates of the signal and pump laser are incommensurable~\cite{comment1}. We obtain a peak centered at zero delay with a linewidth of $6.5 \pm 0.1$~ps as extracted from a Gaussian fit to the data (see fig.~\ref{fig:3}c). We deduce the resolution of the upconversion setup to be $4\pm 0.5$~ps, in agreement with our simulations. In the last step, we feed our setup with a source of weakly subpoissonian light. The emitter consists of a resonantly driven fiber cavity in strong coupling with the excitonic transition of a single quantum well. The structure and the optical setup are extensively described in ref~\cite{Delteil19} and sketched in fig.~\ref{fig:3}d. This source emits weakly antibunched light on top of a bunched background when weakly excited with a negative laser detuning of about half the mode linewidth~\cite{Delteil19, Munoz19}. The result we found is in agreement with the data previously acquired in a pulsed regime~\cite{Delteil19}, unveiling the different timescales associated with the antibunched (visible for $\Delta t < 20$~ps) and bunched (persisting for $\Delta t > 50$~ps) components. Although the minimum of antibunching $g^{(2)}(0) = 0.94 \pm 0.04$ is close to the classical limit, the most striking non-classical property revealed by our data is the violation of the classical inequality $g^{(2)}(0) \geq g^{(2)}(\tau)$~\cite{Loudon}. In our case the value of $g^{(2)}(0)$ lies below the maximum of $g^{(2)}(\Delta t)$ by more than 4~standard deviations, confirming the non-classical nature of the source from a time-resolved approach. We wish to highlight that observation of such a short-lived signature is unfeasible with any other available technique to measure $g^{(2)}(\tau)$.

The resolution of our setup is limited both by the length of the pump pulse and the propagation and conversion saturation in the PPLN crystal. It is however straightforward to further improve it by at least one order of magnitude by using femtosecond pulses together with a shorter crystal (see Supplementary Information). The overall efficiency of $5.3 \cdot 10^{-6}$ is mainly limited by the ratio between the pulse width and the repetition period (see Supplementary Information for detailed contributions to the detection efficiency). It could be increased with a higher repetition rate of the pump laser, as long as the corresponding pulse period is longer than the APD time resolution. Although synchronization between pump and signal is not needed to measure $g^{(2)}(\tau)$, it is also possible to excite the signal emitter in a synchronous way, allowing to measure two-time correlation function $g^{(2)}(t_1,t_2)$ in a dynamical or transient regime~\cite{Mangum13,eloi15}.

To conclude, the technique we developed for photon correlation measurements based on ultrafast light sampling is likely to play an increasingly important role in the measurement of quantum optical properties of ultrafast single photon sources~\cite{Mikkelsen15}, as well as in the investigation of single- or few-photon phenomena in condensed matter occurring at short timescales~\cite{Snijders18,Tighineanu16,Delley17}.

\textbf{Funding.}
Swiss National Science Foundation (SNSF) through a DACH project 200021E-158569-1; SNSF National Centre of Competence in Research - Quantum Science and Technology (NCCR QSIT); ERC Advanced investigator grant (POLTDES).

\textbf{Acknowledgment.}
The Authors gratefully acknowledge many enlightening discussions with K.~Srinivasan. They also thank A.~Schade, C.~Schneider and S.~H\"ofling for the growth of the sample used to obtain the data depicted in Fig.~\ref{fig:3}d.

\bibliography{biblio}

\begin{thebibliography}{26}
\bibitem[{com()}]{comment1}
\bibinfo{note}{In this particular case we do not normalize by the average of
  coincidences from non-zero period difference, due to the vanishing
  coincidence rate when the pump and signal lasers have slightly different
  repetition rates. Instead we divide the same-period coincidence rate by the
  square of the count rate and normalize the peak to 1.}
\end{thebibliography}

\end{document}